\documentclass[aps,prb,twocolumn,amsmath,amssymb]{revtex4}
\usepackage{dcolumn}
\usepackage{bm}
\usepackage{graphics}
\usepackage{graphicx}
\usepackage{color}
\usepackage{epsfig}
\usepackage{amssymb}
\begin{document}
\title{Relating field-induced shift in transition temperature to the kinetics of coexisting phases in magnetic shape memory alloys}
\author{A. Banerjee\footnote{Corresponding author: Email: alok@csr.res.in, Tel.: +91 731 2463913; fax: +91 731 2462294}, S. Dash, Archana Lakhani and P. Chaddah}
\affiliation{UGC-DAE Consortium for Scientific Research, University Campus, Khandwa Road, Indore-452001, Madhya Pradesh, India.}
\author{X. Chen and R. V. Ramanujan}
\affiliation{School of Materials Science and Engineering, Nanyang Technological University, N4.1-01-18, 50 Nanyang Avenue, Singapore 639798.}
\begin{abstract}
In a magnetic shape memory alloy system, we vary composition following phenomenological arguments to tune macroscopic properties. We achieve significantly higher shift in austenite to martensitic phase transition temperature with magnetic field. This enhancement is accompanied by significant broadening of the transition and by field-induced arrest of kinetics, both of which are related to the dynamics of the coexisting phases. This reveals hitherto unknown interrelationship between different length-scales. This may serve as an effective route for comprehensive understanding of similar multicomponent systems which show considerable variation in physical properties by minor change in microscopic parameters. 
\end{abstract}

\maketitle
\section{Introduction}
Interesting physical properties of Heusler alloys have attracted attention for many decades. Recently, a class of such alloys with general formula X$_{2}$MnZ (X = Ni, Co; Z = Sn, Ga, In) has become a subject of intense study because of the effects like magnetic shape memory\cite{kainuma}, magnetocaloric properties\cite{krenke}, half-metallic behavior\cite{sakuraba}, recently discovered unusual exchange bias\cite{wang} etc. Crystallographic structure of these alloys can be constructed from an arrangement of four interpenetrating fcc sublattices, shifted towards body diagonal and the chemical disorder within these sublattices causes drastic changes in the physical properties\cite{brown,li}. Immense importance of disorder becomes apparent in the magnetic shape memory alloys of type Ni$_{2}$Mn$_{1+a}$Z$_{1-a}$ which show wide variation in parameters implying intriguing physics\cite{bose}. The first-order field-temperature (H-T) induced magnetic transition from austenite to martensitic phase is accompanied by a sharp fall in magnetization, which is significantly influenced by the disorder in the Ni sites and the Mn nonstoichiometry. For example, chemical disorder introduced by Co doping in the Ni site of Ni-Mn-In alloy causes colossal stress in magnetic field\cite{kainuma}. It was shown recently that magnetic moments on two Mn sites changes their orientation from being antiparallel to parallel on Co doping in similar alloy\cite{umetsu1}. However,the relationship between these two effects is not clear. In addition, it is shown in NiCoMnIn alloy the martensitic transformation is interrupted while cooling in field resulting in the presence of a kinetically arrested (KA) austenite phase fraction at low temperature\cite{ito,umetsu2}.  Drastic changes related to the first-order phase transition (FOPT) and the KA of the transformation process are considered to bring about tunable coexistence of magnetic phases over a wide H-T space in materials of current interest like CMR manganites\cite{alok1,chaddah}, magnetocaloric materials\cite{roy1}, magnetic shape memory alloys\cite{ito,sharma,chatterjee,llamaKA,umetsu3} and electro-magnetic multiferroics\cite{choi}. 
 
Hence, there is an urgent need to understand various aspects of the first-order austenite to martensitic transformation process and their relation to the chemical composition and disorder. As such these transitions are broad because of the distribution of ``coherence volume" across the sample arising from the hindrance created by the disorder to the divergence of the correlation length, resulting in a distribution of transition temperatures \cite{imry}. Magnetic field dependence of the first-order martensitic to austenite transition temperatures (T$_{C}$) is another important aspect, and a large dT$_{C}$/dH is shown to accompany a huge stress as mentioned above\cite{kainuma}. However, proper identification of T$_{C}$ is very important for such systems having broad FOPT and signature of field induced KA. Recently, it is shown that KA does not interfere with the superheating spinodal (T**) and does not cause any anomaly related to T** \cite{dash}. Hence, in this study we have used the reverse martensitic to austenite transition to define T$_{C}$ as the temperature of the peak in magnetization measured while warming in different fields after cooling in zero field. 

For a broad FOPT, Clausius-Clapeyron relation can be invoked in a limited sense as
\begin{equation}
\frac{dT_{C}}{dH} \cong -\frac{\Delta M}{\Delta S}
\end{equation} 
  where $\Delta$M and $\Delta$S are the changes in magnetization and entropy of the two phases across the broad transition. Relating these thermodynamic parameters to the microscopics and a priori design of a material appears to be a nontrivial task because it is difficult to quantify parameters like ``relevant disorder" for the system and its consequences on different material properties\cite{umetsu2}. However, it may be possible to use phenomenological reasoning to tune some important parameters and unveil their hitherto hidden relationship and obtain a correlation with far reaching consequences. 
  
We initiated this study with Ni$_{50}$Mn$_{35}$Sn$_{15}$ (NMS) and found a small dT$_{C}$/dH, $\Delta$M as well as magnetization in the austenite phase. Following phenomenological arguments we prepared Ni$_{45}$Co$_{5}$Mn$_{50-A}$Sn$_{A}$ (A = 11-15) series and found that the sample Ni$_{45}$Co$_{5}$Mn$_{38}$Sn$_{12}$(Sn-12) has very high dT$_{C}$/dH. However, the first-order austenite to martensitic transformation in Sn-12 sample is much broader compared to other samples of the series. Moreover, this sample shows anomalies related to H-T induced KA of the austenite to martensitic transformation process giving rise to tunable coexistence of magnetic phases down to the lowest temperature as shown for CMR manganite\cite{alok1}. Thus this study reveals a correlation between the broadening of transition and dT$_{C}$/dH and physics at mesoscopic length-scale related to the coexistence of the competing phases because of KA. The ensuing correlation discovered here may  have far-reaching consequences and bring out commonality between diverse systems.

\section{Experimental Details}
The samples are prepared by arc melting of high purity elements under high purity argon atmosphere in water-cooled copper hearth. The button was re-melted several times to ensure homogeneity. The buttons were then used to prepare ribbons by melt spinning in an inert atmosphere at 60 rpm wheel speed. The composition of the melt-spun alloy was determined by an energy dispersive X-ray spectrometer (EDXS) attached to a scanning electron microscope (SEM) (JEOL, JEM-6360).  The crystal structure was investigated by a Bruker (D8 Advance) X-ray diffractometer (XRD) and Rigaku (Rotaflex RTC 300 RC) diffractometer attached to 18 kW rotating anode with Cu K$\alpha$ radiation. Rietveld refinement of XRD data confirmed that the samples are single phase, without any detectable impurity. Magnetization was measured with commercial 14 Tesla VSM (PPMS) made by Quantum Design. 

\section{Results and Discussions}
We have chosen melt-spun ribbon samples for the present investigation because unlike bulk polycrystalline sample they do not need long and repeated annealing to get homogeneous single phase\cite{llama}. Moreover, ribbons are more useful for technological applications\cite{aliev}. The Ni$_{50}$Mn$_{35}$Sn$_{15}$ ribbon sample shows a first order austenite to martensitic transition as a function of temperature. The transition temperatures for austenite to martensitic and martensitic to austenite phases match well with the bulk polycrystalline sample of same composition\cite{chatterjee}. Figure 1a shows the magnetization (M) as a function of temperature (T) while warming in different H after cooling in zero field (ZFC). It is evident that there is minor shift in transition with H, and if we designate the temperature of the peak magnetization as T$_{C}$ then it shifts by $\sim$1.6K/Tesla. It may be noted that the bulk polycrystalline sample has also shown a similar small shift in T$_{C}$  ($<$ 2K/Tesla) with H\cite{chatterjee}.

In the Ni$_{45}$Co$_{5}$Mn$_{50-A}$Sn$_{A}$ (A = 11-15) series of samples it is found that M decreases with increase in Mn and Sn-12 sample shows a higher-M austenite to lower-M martensitic transition with decrease in temperature as shown by the ZFC data in Figure 1b. With further increase in Mn, the Ni$_{45}$Co$_{5}$Mn$_{39}$Sn$_{11}$ (Sn-11) sample shows decrease in M and higher-M austenite to lower-M martensitic transition shifts to higher temperature depicted by ZFC data in Figure 1c. It is evident that while Sn-11 sample shows small shift in transition with H  ($<$ 2K/Tesla) the shift in T$_{C}$ with H for Sn-12 is relatively huge ($>$ 6K/Tesla). This aspect becomes visibly clear in Figure 1d where the temperature of the peak magnetization (T$_{C}$) is plotted as a function of measurement field. Significantly higher dT$_{C}$/dH of Sn-12 sample compared to even the recently discovered Co doped Ni-Mn-In alloy ($\sim$4K/Tesla), which has shown colossal stress in magnetic field\cite{kainuma}, indicates the possibility of superior field induced property.
  
Justification for such high dT$_{C}$/dH does not appear automatically from Eq.1, because, Figure 1 shows that $\Delta$M for Sn-11 and Sn-12 samples are similar and chemical disorder in both the samples are of comparable level. However, the broadening of the FOPT, which is related to disorder, appears to be very different for these two samples. We estimate the width of transition ($\Delta$T) from the data of Figure 1 by taking the temperature difference between temperature of peak magnetization and the temperature at which the magnetization drops by 75$\%$ of the difference between its peak value and the value at 10K. For Sn-11 and NMS samples, $\Delta$T varies from 7K to 12K for field range of 1-12 Tesla. Whereas for Sn-12 sample $\Delta$ T is 17K at 1 Tesla and monotonically increases to 42K at 12 Tesla. This enormous increase in the broadening of transition for Sn-12 compared to Sn-11 having similar level of substitutional chemical disorder as well as Sn-11 and NMS having similar broadening in spite of very different level of chemical disorder is rather intriguing. This does not imply contradiction with the classical work of Imry $\&$ Wortis because we do not have proper understanding about the ``relevant disorder" for the systems under investigation, which broadens the FOPT. Nevertheless, there is a concern, because it is proposed that very excessive disorder can cause a FOPT to become continuous. Further, it is recently shown in some CMR-manganites that the first-order magnetic transition becomes second-order at high-H or pressure\cite{demko,sarkar}. Thus it becomes imperative to investigate the nature of transition for the Sn-12 in the T-H range across the T$_{C}$-H line depicted in Figure 1d. 

It is important to ascertain if the transition as a function of H is first-order for Sn-12 sample. We take recourse to the protocol demonstrated for intermetallic Mn$_{2}$Sb to confirm the field induced FOPT\cite{pallavi1,pallavi2}. We measure isothermal M-H curves after reaching the respective measurement temperatures (T$_m$) by two different paths; once by directly cooling from 350K in zero field and second time by cooling from 350K to 10K and then heating to T$_m$, all in zero field. Since the temperature range of dT$_{C}$/dH line for Sn-12 is from $\sim$144K to 210K, we measure isothermal M-H encompassing this range and show isothermal M-H curves for a few representative temperatures and for field cycle of 0-14-0-(-14)-0-14 Tesla in Figure 2. The panels a-c of Figure 2 are the M-H curves after cooling in zero field from 350K to the respective temperatures (T$_m$) and panels d-f are the M-H curves at the same temperatures T$_m$ reached by heating from 10K, i.e. by traversing the path 350K-10K-T$_m$ in zero field. At 215K the system is above its superheating spinodal and is in stable high-M austenite phase. Hence, the M-H curves obtained by reaching T$_m$ while cooling and while heating are identical. As T$_m$ decreases the system enters the metastable region between the superheating and supercooling limits. In this temperature region the state of the system in zero field is path dependent because while cooling it may undergo partial conversion to low-M martensitic state but when it reaches T$_m$ by heating it has already undergone full conversion to low-M martensitic phase while reaching 10K. Thus, in the heating curves the system starts with a lower magnetization value and the initial field increasing cycle (virgin curve) starts at a lower value and remain distinctly below the envelope M-H curve. However, after causing the field induced transition when field is reduced to zero, it does not fully convert back to low-M martensitic phase initial state because T$_m$ is above the supercooling limit and the envelope curve becomes symmetric. Whereas, the cooling curves do not show such apparent anomaly of the virgin M-H. This is a confirmatory test that the field induced transition is of first-order as explained and also proved through bulk as well as mesoscopic measurements in Refs 25 and 26.   

However, many systems that show signatures of canonical field induced FOPT of the ZFC state, often show anomalies related to KA of the transformation when cooled in non-zero fields. Figure 3a shows the M-T curve for Sn-12 while cooling and then heating in different fields. It is noteworthy that the thermal hysteresis, which is considered to be the signature of temperature induced FOPT, reduces with the increase in measurement field and almost vanishes at 9 Tesla. Inset of Figure 3b shows the M-H curve at 5K after cooling in zero field. ZFC state is low-M martensitic phase and the M-H does not show any field induced transition in the accessible field range. It behaves like a soft ferromagnet having technical saturation above $\sim$1 Tesla with a saturation magnetization less than 80 emu/g. On the contrary, the M-T curves shown in Fig. 3a has much higher magnetization at 5K for curves shown for 3 Tesla or above indicating the presence of a fraction of high-M austenite phase. The progressive increase in M at 5K with the increase in H indicates decrease in the fraction of martensitic phase at low-T. Since austenite phase is also a soft-FM as is evident from the M-H of Fig. 2 (a) and (d), the M values at low-T for H $\geq$ 3 Tesla directly reflect the phase fractions. Thus the first-order transformation kinetics of austenite to martensitic phase with decrease in temperature gets progressively inhibited by the increase in cooling field. At 9 Tesla the FOPT is completely halted and the system approaches low-T with almost fully arrested austenite phase having a saturation magnetization of $\approx$ 132 emu/g. However, it may be noted that the NMS sample, which shows FOPT in the same temperature range, does not show the collapse of the thermal hysteresis even in 9 Tesla (not shown here).  

To confirm the field induced KA of FOPT from austenite to martensitic phase, we follow the well established measurement protocol CHUF (cooling and heating in unequal fields)\cite{alok1,alok2,roy2}. Figure 3b shows a representative measurement following CHUF protocol where magnetization is measured in 4 Tesla while warming after cooling in different fields. While cooling in higher fields, larger fraction of austenite is accrued which remains as a  metastable arrested phase at low-T. If the measuring (warming) field is lower, then this metastable austenite will dearrest to the stable martensitic phase at low-T while warming and the magnetization will show a sharp fall. This will be followed by usual conversion of martensitic to austenite phase at higher temperature showing a double (re-entrant) transition. On the contrary, if the measuring (warming) field is higher than the cooling field then the sharp fall corresponding to dearrest will not occur. The smaller fraction of the arrested austenite phase accrued while cooling in lower field cannot dearrest while warming in higher field because as shown earlier, in this case, the arrested thermodynamically unstable phase has already become metastable in higher field\cite{alok1,chaddah}. Thus there will not be any double transition, only usual conversion of martensitic to austenite will occur at high-T corresponding to superheating limit. In a recent study on a multiferroic material, it has been re-emphasised that the competition of coexisting phases arising from the kinetic arrest and dearrest phenomena is responsible for both physical properties and their control by external parameters for technological applications\cite{choi}. 

However, in spite of recent flurry of activities related to the KA of FOPT, various aspects related to it has remained rather intriguing. For example, recently evidence of KA is given for a NiCoMnSn system and it is asserted that the temperature of the KA is about 100K and almost field independent as identified from the merger of heating and cooling curves\cite{umetsu3}. Though the present system (Sn-12) also show that the merger of heating and cooling curves takes place almost around the same temperature of 60K for all fields(Fig. 3a), the dearrest of the arrested austenite phase fractions starts at different temperatures, varying from 10K to 60K, for different cooling fields as depicted in the Fig. 3b by the sharp fall in M while warming in 4T field. This is an unambiguous evidence of the field dependence of the KA temperature (see Fig.1 of Ref. \cite{kran}). The KA seen here is similar to that studied in various other materials \cite{alok1, chaddah, roy1, dash,pallavi2, alok2, kran} and underscores the generality of the hindered kinetics of a low-temperature first-order transformation process in magnetic systems.       

\section{Conclusions}
In conclusion, for magnetic shape memory alloy, we unveil a relationship between the bulk properties like broadening of a FOPT and its shift with field, to the phenomenon of kinetic arrest that essentially involves dynamics at the mesoscopic length scales. Based on phenomenological arguments involving the microscopic properties, we could tune the material properties. We show that when the FOPT is broad, dT$_{C}$/dH is significantly higher and also there is a field-induced KA resulting in tuneable coexistence of the high-T austenite phase down to the lowest temperatures. There is a growing awareness of the need to understand the dynamics and thermodynamics of coexisting phases at the mesoscopic length scale and relate it to the length scales at the two extremes. However, there are various gaps in the present understanding, which forbids a direct and general connectivity from microscopic to macroscopic length-scales.  The correlation discovered here involving different length scales can provide important pathway towards a comprehensive understanding.

\begin{figure*}
	\centering
		\includegraphics[width=9cm]{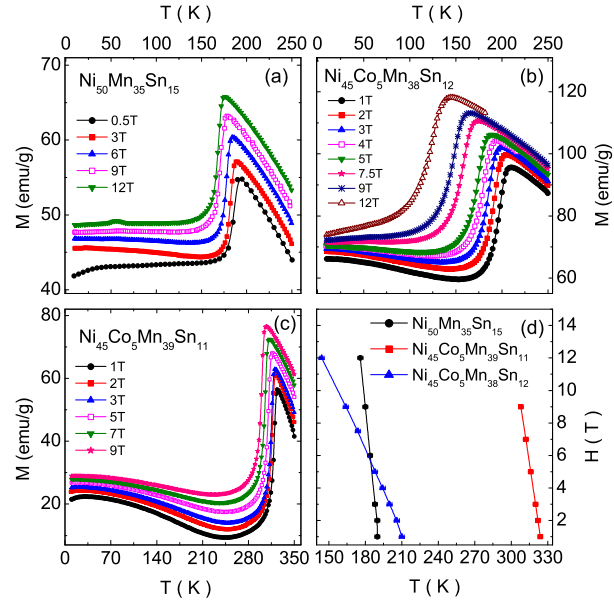}
		\caption{(Color online) Shows the results of magnetization (M) as a function of temperature (T) in different external magnetic field (H) while warming after cooling to 5K from 350K in zero-field (ZFC). ZFC M-T for Ni$_{50}$Mn$_{35}$Sn$_{15}$, Ni$_{45}$Co$_{5}$Mn$_{38}$Sn$_{12}$ and Ni$_{45}$Co$_{5}$Mn$_{39}$Sn$_{11}$ samples are shown in (a), (b) and (c) respectively. The temperature of the peak magnetization as a function of measuring H taken from (a)-(c) for the respective samples and is shown in (d).}
\label{fig:fig1}
\end{figure*}

\begin{figure*}
	\centering
		\includegraphics[width=9cm]{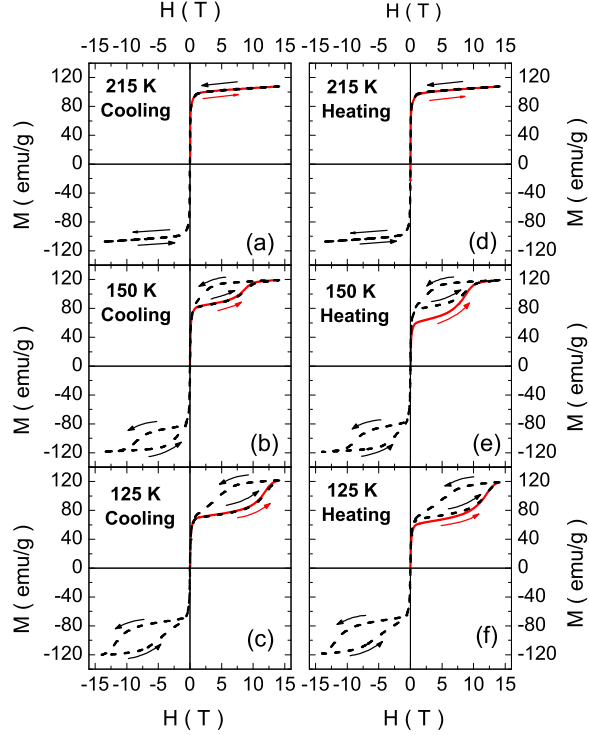}
		\caption{(Color online) Isothermal M-H curves for Ni$_{45}$Co$_{5}$Mn$_{38}$Sn$_{12}$ sample. The measurement temperatures (T$_m$) are reached by two different paths. (a)-(c) show the M-H when the respective T$_m$ is reached directly by cooling from 350K. (d)-(f) shows the M-H when the respective T$_m$ is reached after cooling from 350K to 10K and then heating from 10K to T$_m$. Initial field increasing cycle (virgin curve) is plotted as a continuous line (red) and the subsequent field excursion (envelope curve) is plotted as dotted line (black).}
\label{fig:fig2}
\end{figure*}

\begin{figure*}
	\centering
		\includegraphics[width=9cm]{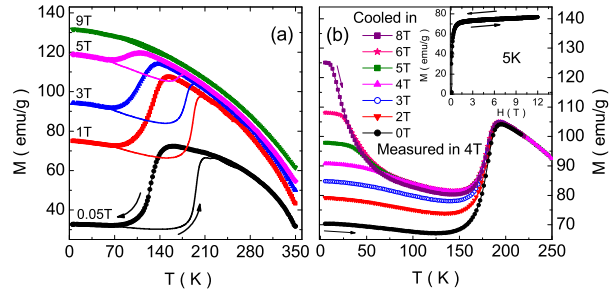}
		\caption{(Color online) M as a function of T for Ni$_{45}$Co$_{5}$Mn$_{38}$Sn$_{12}$ sample. (a) Thermal hysteresis in M related to the first-order transition in various measurement fields. The measuring fields are given alongside the respective curves. (b) M while warming in 4 Tesla after cooling in different fields following CHUF protocol. The fields are isothermally changed to 4 Tesla at 5K after cooling in different fields.}
\label{fig:fig3}
\end{figure*}
\end{document}